\documentclass[conference]{IEEEtran}
\IEEEoverridecommandlockouts
\usepackage{ifpdf}
\usepackage{cite}
\usepackage[pdftex]{graphicx}
\usepackage{amsmath}
\usepackage{algorithmic}
\usepackage{array}
\usepackage{subfigure}
\usepackage{caption}
\usepackage[caption=false,font=footnotesize]{subfig}
\usepackage{fixltx2e}
\usepackage{stfloats}
\usepackage{url}
\usepackage{amsthm}
\usepackage{amsmath}
\usepackage{amsfonts}
\usepackage{amssymb}
\usepackage[T1]{fontenc} 
\usepackage{amsmath}
\usepackage[cmintegrals]{newtxmath}
\usepackage{bm} 
\usepackage[all]{xy}
\usepackage{enumitem}
\usepackage{float}
\usepackage{amsmath}
\usepackage[usenames, dvipsnames]{color}
\usepackage{ifpdf}
\usepackage{cite}
\usepackage[pdftex]{graphicx}
\usepackage{amsmath}
\usepackage{algorithmic}
\usepackage{array}
\usepackage{mathtools}
\DeclarePairedDelimiter{\abs}{\lvert}{\rvert}

\usepackage{caption}
\usepackage[caption=false,font=footnotesize]{subfig}
\usepackage{fixltx2e}
\usepackage{stfloats}
\usepackage{url}
\usepackage{mathtools}
\usepackage{amsthm}
\usepackage{amsmath}
\usepackage{amsfonts}
\usepackage{amssymb}
\usepackage[T1]{fontenc} 
\usepackage{amsmath}
\usepackage[cmintegrals]{newtxmath}
\usepackage{bm} 
\usepackage[all]{xy}
\usepackage{enumitem}
\usepackage{float}
\usepackage{amsmath}
\usepackage[dvipsnames]{xcolor}
\usepackage{multirow}
\usepackage{subfig}
\usepackage{multicol}
\usepackage{graphicx}
\usepackage[caption=false]{subfig}
\usepackage{amssymb}
\usepackage{amsmath}
\usepackage{commath}
\usepackage{graphicx,bm}
\usepackage{verbatim}

\usepackage{diagbox}

\theoremstyle{definition}

\newtheorem{thm}{Theorem}

\newtheorem{lem}{Lemma}

\newtheorem{define}{Definition}
\newcommand{\cov}{\textrm{Cov}}

\newcommand{\no}{\nonumber}

\begin{document}
\title{Privacy against Statistical Matching:\\  Inter-User Correlation}

\author{\IEEEauthorblockN{Nazanin Takbiri}
\IEEEauthorblockA{Electrical and\\Computer Engineering\\
UMass-Amherst\\
ntakbiri@umass.edu}
\and
\IEEEauthorblockN{Amir Houmansadr}
\IEEEauthorblockA{Information and \\Computer Sciences\\
UMass-Amherst \\
amir@cs.umass.edu}
\and
\IEEEauthorblockN{Dennis L. Goeckel}
\IEEEauthorblockA{Electrical and\\Computer Engineering\\
UMass-Amherst\\
goeckel@ecs.umass.edu}
\and
\IEEEauthorblockN{Hossein Pishro-Nik}
\IEEEauthorblockA{Electrical and\\Computer Engineering\\
UMass-Amherst\\
pishro@ecs.umass.edu\thanks{This work was supported by National Science Foundation under grants CCF--1421957 and CNS-1739462.}}
}

\maketitle

\begin{abstract}
Modern applications significantly enhance user experience by adapting to each user's individual condition and/or preferences. 
While this adaptation can greatly improve utility or be essential for the application to work (e.g., for ride-sharing applications), the exposure of user data to the application presents a significant privacy threat to the users, even when the traces are anonymized, since the statistical matching of an anonymized trace to prior user behavior can identify a user and their habits.  Because of the current and growing algorithmic and computational capabilities of adversaries, provable privacy guarantees as a function of the degree of anonymization and obfuscation of the traces are necessary.  Our previous work 
has established the requirements on anonymization and obfuscation in the case that data traces are independent between users.  However, the data traces of different users will be dependent in many applications, and an adversary can potentially exploit such.  In this paper, we consider the impact of correlation between user traces on their privacy.  First, we demonstrate that the adversary can readily identify the association graph, revealing which user data traces are correlated.  Next, we demonstrate that the adversary can use this association graph to break user privacy with significantly shorter traces than in the case when traces are independent between users, and that independent obfuscation of the data traces is often insufficient to remedy such. Finally, we discuss how the users can employ dependency in their obfuscation to improve their privacy.  

\end{abstract}

\begin{IEEEkeywords}
Internet of Things (IoT),  Privacy-Protection Mechanisms (PPM), Information Theoretic Privacy, Anonymization, Obfuscation, Inter-User Correlation.
\end{IEEEkeywords}


\section{Introduction}
\label{intro}

Many modern applications exploit a user's characteristics, both their past choices and present state, to enhance user experience.  For example, emerging Internet of Things (IoT) applications include smart homes, health care, and connected vehicles that will smartly tune their response to a given user, such as a connected vehicle application optimizing a route based on the current location of the vehicle and traffic conditions.  Such applications require that a user provide potentially sensitive data; hence, questions arise about how much user privacy is compromised.  Even if user data traces are anonymized, statistical matching of the current data trace with prior user behavior can identify the user and their characteristics \cite{Naini2016,matching}.  And studies have indicated that privacy concerns could significantly hamper the penetration of IoT applications \cite{ iotCastle,0Quest2016,3ukil2014iot}.

Our previous work has introduced the notion of ``perfect privacy''~\cite{tifs2016}. In particular, with rapid advances in algorithms and computation, information-theoretic guarantees that demonstrate that sensitive information does not leak to a powerful adversary are critical. The work of~\cite{ISIT18-longversion1,nazanin_ISIT2017,tifs2016} considered the degree of user anonymization and data obfuscation required to obtain perfect privacy in the case when the data traces of different users are independent of one another. In that work, we have considered the case of independent and identically distributed (i.i.d.) samples from a given user and the case when there is temporal correlation within the trace of a given user~\cite{ISIT18-longversion1,nazanin_ISIT2017}, as have others under different metrics~\cite{corgeo,corPLP,cordummy,cordiff}.

There are many applications where there is correlation between the traces of different users.  For example, friends tend to travel together or might meet at given places, hence introducing dependency between locations.  However, there is relatively limited work in this area \cite{cordiff8, cordata1, cordiff2, cordiff3, cordiff4}, particularly from a fundamental perspective.  Hence, here we investigate what the metrics of  ~\cite{ISIT18-longversion1,nazanin_ISIT2017,tifs2016} require in terms of user anonymization and data obfuscation to preserve privacy in the case of correlation between the data traces of different users.

We model dependence between user traces with an association graph, where an edge between the vertices corresponding to a pair of users indicates dependency between their data traces. We first demonstrate that the adversary can readily determine this association graph. Armed with this association graph, the adversary can attempt to identify the users, and we show that this provides the adversary with a significant advantage versus the case when the data traces of different users are independent of one another.  This suggests that, unless additional countermeasures are employed, the results of \cite{ISIT18-longversion1,nazanin_ISIT2017,tifs2016,ciss2017,ciss2018} for independent traces are overly optimistic when user traces are correlated.   We next consider countermeasures.  First, we demonstrate that adding independent obfuscation to user data samples is often ineffective in improving the users' privacy.  Finally, we demonstrate that, if users with correlated traces can jointly design their obfuscation, user privacy can be significantly improved.


More references, additional discussion, and proofs of main results are provided in the long version of the paper \cite{ISIT18-longversion}.



\section{Framework}
\label{sec:framework}
\begin{figure}[h]
	\centering
	\includegraphics[width = 1\linewidth]{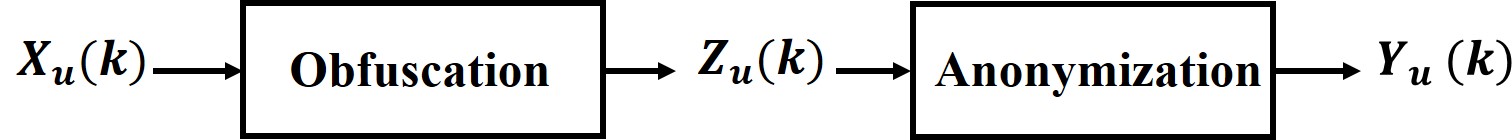}
	\caption{Notation for sequences after applying obfuscation and anonymization to users' data samples.}
	\label{fig:xyz}
\end{figure}

We employ a similar framework to~\cite{tifs2016,nazanin_ISIT2017}.  The system has $n$ users, and $X_u(k)$ is the sample of the data of user $u$ at time $k$. Our main goal is protecting $X_u(k)$ from a strong adversary $(\mathcal{A})$ who has full knowledge of the (unique) marginal probability distribution function of the data samples for each user based on previous observations or other resources. In order to achieve data privacy of users, both anonymization and obfuscation techniques can be used as shown in Figure \ref{fig:xyz}. In Figure \ref{fig:xyz}, $Z_u(k)$ shows the (reported) sample of the data of user $u$ at time $k$ after applying obfuscation, and $Y_u(k)$ shows the (reported) sample of the data of user $u$ at time $k$ after applying anonymization.  Let $m=m(n)$ be the number of data points after which the pseudonyms of users are changed in the anonymization.  To break the anonymization, the adversary tries to estimate $X_u(k)$, $k=1, 2, \cdots, m$, from $m$ observations per user by matching the sequence of observations to the known statistical characteristics of the users. 

Let $\textbf{X}_u$ be the $m \times 1$ vector containing the samples of the data of user $u$, and $\textbf{X}$ be the $m \times n$ matrix with $u^{th}$ column equal to $\textbf{X}_u$;
\[\textbf{X}_u =\left [
X_u(1), X_u(2) , \cdots,X_u(m)\right]^T \]
\[ \textbf{X} =\left[\textbf{X}_{1}, \textbf{X}_{2}, \cdots,  \textbf{X}_{n}\right].
\]

\textit{Data Samples Model:}
We assume users' data samples can have $r$ possibilities $(0, 1, \cdots, r-1)$. Thus, according to a user-specific probability distribution, $X_u(k)$ is equal to a value in $\left\{0,1, \cdots, r-1 \right\}$ at any time, and, per above, these user-specific probability distributions are known to the adversary $(\mathcal{A})$ and form the basis upon which he performs (statistical) matching.

\textit{Association Graph:}
An association graph or dependency graph is an undirected graph representing dependencies of users with each other. Let $G(V,F)$ denote the association graph with set of nodes $V$, $(|V|=n)$, and set of edges $F$. Two vertices (users) are connected if their data sets are dependent. More specifically,
\begin{itemize}
	\item $(u,{u'}) \in F$ if $I(X_u(k); X_{{u'}}(k))>0,$
	\item $(u,{u'}) \notin F$ if $I (X_u(k);X_{{u'}}(k))=0,$
\end{itemize}
where $I(X_u(k); X_{u'}(k))$ is the mutual information between the $k^{th}$ data sample of user $u$ and user ${u'}$.

\textit{Obfuscation Model:}  Obfuscation perturbs the users' data samples. Each user has only limited knowledge of the characteristics of the overall population, so usually a simple distributed method in which the samples of the data of each user are reported with error with a certain probability is employed. Note that this probability itself is generated randomly for each user. Let ${\textbf{Z}}_u$ be the vector which contains
the obfuscated versions of user $u$'s data sample, and ${\textbf{Z}}$ be the collection of ${\textbf{Z}}_u$ for all users,

%

\textit{Anonymization Model:} In the anonymization technique, the identity of the users is perturbed. Anonymization is modeled by a random permutation $\Pi$ on the set of $n$ users. Let ${\textbf{Y}}_u$ be the vector which contains the anonymized version of ${\textbf{Z}}_u$ (the obfuscated version of data of user $u$), and ${\textbf{Y}}$ is the collection of ${\textbf{Y}}_u$ for all users, thus ${\textbf{Y}}_{u} ={\textbf{Y}}_{\Pi^{-1}(u)}$ and ${\textbf{Y}}_{\Pi(u)} = {\textbf{Y}}_{u}$.

\textit{Adversary Model:} We assume the adversary has full knowledge of the marginal probability distribution function of each of the users on $\{0,1,\ldots,r-1\}$.  The adversary also knows the anonymization mechanism, but does not know the realization of the random permutation. The adversary knows the obfuscation mechanism, but does not know the realization of the noise parameters. Also, the adversary knows the association graph $G(V,F)$, but does not necessarily know the exact nature of dependency. That is, while the adversary knows the marginal distributions $X_u(k)$ as well as which pairs of users have strictly positive mutual information, he might not know the joint distributions or even the values of mutual informations $I(X_u(k); X_{u'}(k))$. It is critical to note
that we assume the adversary does not have any auxiliary information or side information about users' data.  

\begin{define}
User $u$ has \emph{perfect privacy} \cite{tifs2016} if and only if
\begin{align}
\no  \forall k\in \mathbb{N}, \ \ \ \lim\limits_{n\rightarrow \infty} I \left(X_u(k);{{\textbf{Y}}}\right) =0,
\end{align}
where $I\left(X_u(k);{{\textbf{Y}}}\right)$ denotes the mutual information between a sample of the data of user $u$ at time $k$ and the collection of the adversary's observations for all of the users.
\end{define}

\begin{define}
User $u$ has \emph{no privacy} \cite{nazanin_ISIT2017} if and only if there exists an algorithm for the adversary to estimate $X_u(k)$ perfectly as $n$ goes to infinity. In other words, as $n \rightarrow \infty$,
\begin{align}
\no  \forall k\in \mathbb{N}, \ \ \ P_e(u,k)\triangleq P\left(\widetilde{X_u(k)} \neq X_u(k)\right)\rightarrow 0,
\end{align}
where $\widetilde{X_u(k)}$ is the estimated value of $X_u(k)$ by the adversary.
\end{define}

\section{Impact of Correlation Between Users on Privacy Using Anonymization}
\label{anon}
In this section, we consider only anonymization and thus the obfuscation block in Figure \ref{fig:xyz} is not present.


\subsection{$i.i.d.$ Two-State Model}
\label{iidtwo}

There is potentially correlation between the data of different users, but we assume here that the sequence of data for any individual user is $i.i.d.$. The $i.i.d.$ model would apply directly to data that is sampled at a low rate. In addition, understanding the i.i.d. case can also be considered the first step toward understanding the more complicated case where there is dependency. 

We first consider the $i.i.d.$ two-state $(r=2)$ case, where the sample of the data of user $u$ at any time is a Bernoulli random variable with parameter $p_u$, which we define as the probability of user $u$ being at state $1$. Thus, \[X_u(k) \sim Bernoulli \left(p_u\right).\]
The parameters $p_u$, $u=1, 2, \cdots, n$ are drawn independently from a continuous density function, $f_P(p_u)$, on the $(0,1)$ interval. The density $f_P(p_u)$ might be unknown, so all that is assumed here is that such a density exists. From the results of the paper, it will be evident that knowing or not knowing $f_P(p_u)$ does not change the results asymptotically.  Further, we assume there are $\delta_1, \delta_2>0$ such that\footnote{The condition $\delta_1<f_P(p_u) <\delta_2$ is not actually necessary for the results and can be relaxed; however, we keep it here to avoid unnecessary technicalities.}:
\begin{equation}
\no\begin{cases}
    \delta_1\leq f_P(p_u) \leq\delta_2, & p_u \in (0,1).\\
    f_P(p_u)=0, &  p _u\notin (0,1).
\end{cases}
\end{equation}
The adversary knows the values of $p_u$, $u=1, 2, \cdots, n$, and uses this knowledge to match the observed traces to the users. We will use capital letters (i.e., $P_u$) when we are referring to the random variable, and use lower case (i.e., $p_u$) to refer to the realization of $P_u$.

A vector containing the permutation of those probabilities after anonymization is $\widetilde{\textbf{P}}$ , where $\widetilde{P}_{u} = {{P}}_{\Pi^{-1}(u)}$ and  $\widetilde{P}_{\Pi(u)} = {{P}}_{u}$. As a result, for $u=1,2,..., n$, the distribution of the data symbols for the user with pseudonym $u$ is given by:
\[ {Y}_u(k) \sim Bernoulli \left(\widetilde{P}_u\right) \sim Bernoulli\left(P_{\Pi^{-1}(u)}\right)  .\]

For this case, dependency and correlation of the data samples are equivalent, that is, we can say:
\begin{itemize}
	\item $(u,{u'}) \in F$ if $\rho (X_u(k),X_{u'}(k))=\rho_{u{u'}}>0,$
	\item $(u,{u'}) \notin F$ if $\rho (X_u(k),X_{u'}(k))=\rho_{u{u'}}=0,$
\end{itemize}
where $\rho_{u{u'}}$ is the correlation coefficient between the data of user $u$ and that of user ${u'}$. The adversary knows the association graph $G(V,F)$, but does not necessarily know the correlation coefficient $(\rho_{u{u'}})$ for each specific $(u,{u'}) \in F$.

\begin{figure}
	\centering
	\includegraphics[width=.6\linewidth]{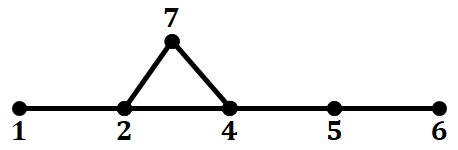}
	\centering
	\caption{One example of an association graph for which the identity map is the only automorphism.}
	\label{fig:identity}
\end{figure}

Critical to compromising the privacy of the users will be the adversary's ability to match empirical correlation properties of the data traces to the known structure of the (ensemble) correlation between users.  First, we show that the adversary can reliably reconstruct the entire association graph for \textit{the anonymized version of the data} (i.e. the observed data traces) with relatively few observations.
\begin{lem}
	\label{lem1}
	Consider a general association graph $G(V,F)$. If the adversary obtains $m=(\log n)^3$ anonymized observations per user, he/she can construct $\widetilde{G}=\widetilde{G}(\widetilde{V}, \widetilde{F})$, where $\widetilde{V}=\{\Pi(u):u \in V\}=V$, such that for all $u, {u'} \in V$; $ (u,{u'})\in F$ iff  $\left(\Pi(u),\Pi({u'})\right)\in \widetilde{F}$. We write this statement as $P(\widetilde{F}=F)\rightarrow 1$.
\end{lem}
The structure of the association graph $(G)$ can leak a significant amount of information.  For example, in Figure \ref{fig:identity}, the identity map is the only automorphism of the association graph $G$.  Thus, it is obvious that the adversary can uniquely identify all of the users if he/she can reconstruct the association graph.

\begin{figure}
	\centering
	\includegraphics[width=.5\linewidth]{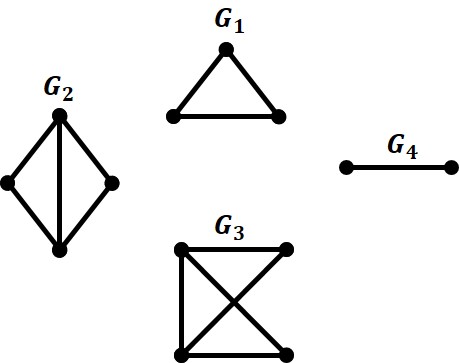}
	\centering
	\caption{Association graph consists of some disjoint subgraphs $(G_j)$, where $G_j$ is a connected graph on $s_j$ vertices.}
	\label{fig:graph}
\end{figure}

To be able to derive further results, we need to make some assumptions on the structure of the association graph. For the rest of the paper, we consider a graph structure shown in Figure \ref{fig:graph}, where the association graph consists of $f$ disjoint subgraphs,
\[G=G_1 \cup G_2 \cup \cdots \cup  G_f,\]
where subgraph $G_j$ is a connected graph on $s_j$ vertices. In particular, each subgraph $G_j$ refers to a group of ``friends'' or ``associates'' such that their data sets are dependent, and we will denote its association graph as $G_j(V_j, F_j)$, where $|V_j|=s_j$.

The following theorem states that if the number of observations per user $(m)$ is significantly larger than $n^{\frac{2}{s}} $ in this two-state model, then the adversary can successfully de-anonymize the users in any group of size $s$.

\begin{thm}\label{two_state_thm}
For the above two-state model, if ${\textbf{Y}}$ is the anonymized version of $\textbf{X}$ as defined above, the size of a group is $s$, and $m=cn^{\frac{2}{s}+\alpha}$, where $\alpha > 0$, then user $1$ has no privacy as $n$ goes to infinity. 
\end{thm}

\textit{Discussion}:
It is insightful to compare this result to Theorem 1 in \cite{tifs2016}, where it is stated that if the users are not correlated, then all users have perfect privacy as long as the number of adversary's observations per user $(m)$ is smaller than $O(n^2)$. Here, Theorem \ref{two_state_thm} states that with much smaller $m$ the adversary can de-anonymize all users. Therefore, we see that correlation can significantly reduce the privacy of users.
$\\$


\subsection{$i.i.d.$ $r$-States Model}
\label{iidr}
Now, assume users' data samples can have $r$ possibilities $\left(0, 1, \cdots, r-1\right)$, and $p_u(i)$ gives the probability of user $u$ having data sample $i$.  We define the vector $\textbf{p}_u$ as
\[\textbf{p}_u= [
p_u(1) ,  p_u(2) ,  \cdots, p_u(r-1) ]^T, \ \ \  \textbf{P} =\left[ \textbf{p}_{1}, \textbf{p}_{2}, \cdots,  \textbf{p}_{n}\right].
\]

We also assume $\textbf{p}_u$'s are drawn independently from some continuous density function, $f_P(\textbf{p}_u)$, which has support on a subset of the $(0,1)^{r-1}$ hypercube. In particular, define the range of the distribution as
\begin{align}
\no  \mathcal{R}_{\textbf{P}} = \{ (x_1, x_2, \cdots, x_{r-1}) &\in (0,1)^{r-1}: \\ \no & x_i > 0 , x_1+x_2+\cdots+x_{r-1} < 1\}.
\end{align}
Then, we assume there are $\delta_1, \delta_2>0$ such that:
\begin{equation}
\begin{cases}
\no    \delta_1\leq f_{\textbf{P}}(\mathbf{p}_u) \leq \delta_2, & \textbf{p}_u \in \mathcal{R}_{\textbf{P}}.\\
    f_{\textbf{P}}(\mathbf{p}_u)=0, &  \textbf{p}_u \notin \mathcal{R}_{\textbf{P}}.
\end{cases}
\end{equation}

\begin{thm}\label{r_state_thm}
For the above $r$-states model, if ${\textbf{Y}}$ is the anonymized version of $\textbf{X}$ as defined above, the size of a group is $s$, and  $m=cn^{\frac{2}{(r-1)s}+\alpha}$, where $\alpha > 0$,
then user $1$ has no privacy as $n$ goes to infinity. 
\end{thm}

\subsection{Markov Chain Model}
\label{markov}
In Sections \ref{iidtwo} and \ref{iidr}, we assumed each user's data patterns was $i.i.d.$; however, in this section users' data patterns are modeled using Markov chains in which each user's data samples are dependent over time. In this model, we again assume there are $r$ possibilities for each data point, i.e., $X_u(k) \in \{0, 1, \cdots, r-1\}$.

More specifically, each user's data is modeled by a Markov chain with $r$ states. It is assumed that the Markov chains of all users have the same structure, but have different transition probabilities. Let $E$ be the set of edges in the assumed transition graph, so $(i, l) \in E$ if there exists an edge from state $i$ to state $l$, meaning that $p_u(i,l)=P\left(X_u(k+1)=l|X_u(k)=i\right)>0$. The transition matrix is a square matrix used to describe the transitions of a Markov chain; thus, different users can have different transition probability matrices. Note for each state $i$, we have
$\sum\limits_{l=1}^{r} p_u(i,l)=1,$
so the adversary can focus on a subset of size $d=|E|-r$ of transition probabilities for recovering the entire transition matrix.
So we have
\[\textbf{p}_u= [
p_u(1), p_u(2), \cdots, p_u(d) ]^T, \ \ \  \textbf{P} =\left[ \textbf{p}_{1}, \textbf{p}_{2}, \cdots,  \textbf{p}_{n}\right].
\]

We also consider $p_u(i)$'s are drawn independently from some continuous density function, $f_P(\textbf{p}_u)$, on the $(0,1)^{|E|-r}$ hypercube. Define the range of the distribution as
\begin{align}
\no  \mathcal{R}_{\textbf{P}} &= \left\{ (x_1, \cdots, x_{d}) \in (0,1)^{d}: x_i > 0 , x_1+x_2+\cdots+x_{d} < 1\right\}.
\end{align}

As before, we assume there are $ \delta_1, \delta_2 >0$, such that
\begin{equation}
\begin{cases}
\no   \delta_1\leq f_{\textbf{P}}(\textbf{p}) \leq \delta_2, & \textbf{p} \in \mathcal{R}_{\textbf{p}}\\
f_{\textbf{P}}(\textbf{p})=0, &  \textbf{p} \notin \mathcal{R}_{\textbf{p}}
\end{cases}
\end{equation}

\begin{thm}\label{markov_thm}
	For an irreducible, aperiodic Markov chain model, if ${\textbf{Y}}$ is the anonymized version of $\textbf{X}$  as defined above, the size of a group is $s$, and $m=cn^{\frac{2}{s(|E|-r)}+\alpha}$, where $\alpha > 0$, then user $1$ has no privacy as $n$ goes to infinity. 
\end{thm}

\section{Privacy using Anonymization and Obfuscation}
\label{obfs}
Here we consider the case when both anonymization and obfuscation techniques are employed, as shown in Figure \ref{fig:xyz}. We assume similar obfuscation to \cite{nazanin_ISIT2017}.

\subsection{$i.i.d.$ Two-State Model}
\label{iidtwo_obfs}
Again, let us start with the $i.i.d.$ two-state model. As before, we assume that $p_u$'s are drawn independently from some continuous density function, $f_P(p_u)$, on the $(0,1)$ interval.

To obfuscate the data samples, for each user $u$ we independently generate a random variable $R_u$ that is uniformly distributed between $0$ and $a_n$. The value of $R_u$ shows the probability that the user's data sample is changed to a different value by obfuscation, and $a_n$ is termed the ``noise level'' of the system.

The effect of the obfuscation is to alter the probability distribution function of each user's data samples in a way that is unknown to the adversary, since it is independent of all past activity of the user, and hence the obfuscation inhibits user identification. For each user, $R_u$ is generated once and is kept constant for the collection of samples of length $m$, thus providing a very low-weight obfuscation algorithm.

The ${Z}_u(k)$'s are $i.i.d.$ with a Bernoulli distribution; thus, 
\[{Y}_u(k) \sim Bernoulli\left(\widehat{p}_u\right),\]
where $\widehat{p}_u$ is is the probability that an obfuscated data sample of user $u$ is equal to one, so
\begin{align}
\no \widehat{p}_u &=p_u(1-R_u)+(1-p_u)R_u \\
\nonumber &= p_u+\left(1-2p_u\right)R_u. \ \
\end{align}
Define the vector $\widehat{\textbf{P}}$, which contains the obfuscated probabilities:
\[\widehat{\textbf{P}} =\left[\widehat{{p}}_1, \widehat{{p}}_2, \cdots, \widehat{{p}}_n\right],\]
and the vector containing the permutation of those probabilities after anonymization as $\widetilde{\textbf{P}}$. As a result,  for $u=1,2,..., n$,
\[ {Y}_u(k) \sim Bernoulli \left(\widetilde{p}_u\right).\]

\begin{thm}\label{two_state_thm_converse_obfs}
	For the above two-state model, if ${\textbf{Z}}$ is the obfuscated version of $\textbf{X}$, and ${\textbf{Y}}$ is the anonymized version of ${\textbf{Z}}$ as defined above, the size of a group is $s$, and
	\begin{itemize}
		\item $m =cn^{\frac{2}{s} +  \alpha}$ for $c>0$ and $\alpha>0$;
		\item $R_u \sim Uniform [0, a_n]$, where $a_n \triangleq c'n^{-\left(\frac{1}{s}+\beta\right)}$ for $c'>0$ and $\beta>\frac{\alpha}{4}$;
	\end{itemize}
	then user $1$ has no privacy as $n$ goes to infinity. 
\end{thm}

\textit{Discussion}:
It is insightful to compare this result to Theorem $1$ in \cite{nazanin_ISIT2017}, which stated that if the users are not correlated, then all users have perfect privacy as long as the number of the adversary's observations per user $(m)$ is smaller than $O(n^2)$ or the noise level $(a_n)$ used to obfuscate the users' data samples is larger than $O(n^{-1})$. Here, Theorem \ref{two_state_thm_converse_obfs} states that with much smaller $m$ or much larger $a_n$ the adversary can de-anonymize and de-obfuscate all users with vanishing error probability. Therefore, we see that correlation can significantly reduce the privacy of users.


\subsection{$i.i.d.$ $r$-States Model and Markov Chain Model}
Similar to sections \ref{iidr} and \ref{markov}, we can extend the results of Section \ref{iidtwo_obfs} to the $r$-states models well as the Markov chain models. The details are provided in the extended version \cite{ISIT18-longversion}.

\section{Achieving Perfect Privacy in the Presence of Correlation}
\label{perfect}
Here, we discuss how we can improve privacy in the presence of correlation. First note that independent obfuscation alone is not sufficient even at a high noise level, because it cannot change the association graph.  Thus, we suggest that associated users collaborate together to increase their privacy. For clarity, we focus on the two-state model with $s_j\leq 2$; thus, there are some users that are connected together and there are also some isolated users.  The asymptotic noise level is defined as the highest probable percentage of data points that are corrupted \cite{ISIT18-longversion} .

\begin{thm}
\label{thm:noise}
For the  two-state model, if ${\textbf{Z}}$ is the obfuscated version of $\textbf{X}$, ${\textbf{Y}}$ is the anonymized version of ${\textbf{Z}}$, and the size of all subgraphs are less than or equal to $2$, there exists an anonymization/obfuscation scheme such that for all $(u,{u'}) \in F$, the asymptotic noise level for users $u$ and ${u'}$ is at most
\[a(u,{u'})= \frac{\abs{\cov (X_u(k), X_{u'}(k))}}{\max \{p_u, p_{u'}, 1-p_u, 1-p_{u'}\}},\]
to achieve perfect privacy for all users.
\end{thm}




\section{Conclusion}
\label{conclusion}

A sophisticated adversary can threaten user privacy by employing statistical matching of user data traces to prior behavior.  Our previous work has considered the requirements on anonymization and obfuscation for ``perfect'' user privacy to be maintained when traces are independent between users.  But traces are rarely independent, as relationships between users establish dependence in their behavior.  We have shown here that such dependence can have a significant impact on user privacy, as the anonymization employed must be significantly increased to preserve perfect privacy, and often no degree of independent obfuscation of the traces can be effective.  We also present preliminary results on dependent obfuscation to preserve user privacy.

\appendices

\bibliographystyle{IEEEtran}
\bibliography{REF}
%
%
%
%

 \end{document}